\documentclass[12pt]{article}

\title{Lorentz Symmetry and Ultra-High-Energy Cosmic Rays.}
\author{M. Toller \thanks{e-mail: toller@iol.it}\\ 
via Malfatti n. 8  \\
I-38100 Trento, Italy}
 
\begin{document} 
\maketitle
                 
\begin{abstract}
We discuss the possibility of explaining the observation of ultra-high-energy cosmic rays with energy above the GZK cutoff, saving the relativity principle and the (possibly deformed) Lorentz symmetry, as proposed recently by several authors. Since it is known that the Lie group structure of the Lorentz group cannot be deformed, we study the deformations (up to isomorphisms) of the mass-shell, considered as an abstract three-dimensional homogeneous space. We find that in the massive case the mass-shell cannot be deformed and in the massless case there are deformations, but their physical interpretation is problematic. The components of the four-momentum are considered as (redundant) coordinates on the abstract mass-shell. Reinterpreting an old result, we note that if the four-momentum is conserved its components must be the usual ones, with linear Lorentz transformation properties. Even if four-momentum is not conserved at high center-of-mass energies, the linearly transforming coordinates can always be used to describe in a convenient way the kinematics of collision processes and they satisfy the GKZ cutoff. We suggest that, if one wants to save the relativity principle, one should look for new physics in the collisions between the ultra-high-energy cosmic rays and the nuclei of the high atmosphere.

\end{abstract}

\newpage

\section{Introduction.}
Shortly after the discovery of the cosmic microwave background (CMB), it has been remarked \cite{ZK,Greisen} that cosmic ray protons may interact with the photons of the CMB and that, if their energy is larger than $5 \times 10^{10} \, GeV$ (the GZK energy cutoff), the center-of-mass energy can be larger than the threshold for pion photoproduction, a reaction with a large cross section. If the protons have an extragalactic origin, as it is suggested by the apparently uniform distribution of their directions, one expects that only protons with an energy smaller than the GZK  cutoff can reach the earth.  However, recent experiments \cite{Takeda} have detected cosmic rays with substantially larger energies (see, however, ref.\ \cite{BW} and references therein).

It has been suggested \cite{ACEMNS,Kifune,CG2,GM,BC,SG,ACP,AC1} that this problem, and possibly other problems in cosmic ray physics, can be solved introducing some deformation of the kinematical laws which govern the propagation and the collision of particles. In particular, it has been proposed, with various theoretical motivations, \cite{CG2,CK,KL}, to modify the dispersion law which connects the energy $p^0$ and the momentum $\vec p$ by assuming 
\begin{equation} \label{Disp}
(\vec p)^2 = h(p^0) \neq (p^0)^2 - m^2, \qquad (c = 1),
\end{equation}
where $p^0$ varies, as usual, on a half line or, possibly, on a bounded interval. The function $h$ contains a new high-energy scale and it has been suggested \cite{ACEMNS,Kifune,ACP,AC1} that this could be one of the few possibly observable consequences of quantum gravity.

The modified dispersion law is rotationally invariant, but not Lorentz invariant. One may assume that it is valid only in some privileged inertial frames (for instance the frames in which the CMB is approximately isotropic), while in other frames it has a different (non rotationally invariant) form. A different assumption \cite{AC4,MS} is that, in agreement with the relativity principle, the dispersion law is the same in all the inertial frames, but the Lorentz transformation properties have to be modified. Sometimes, this proposal, is called ``doubly special relativity'', because the symmetry transformations respect both the velocity of light and the new energy scale. In the last two years, it has stimulated a wide discussion \cite{KG,BACKG,AC2,ACBD,KGN,KGN2,LN,JV,RS,LN2,KG2,Bruno,Granik,AC5,MS2,AC3,LN3,LN4,KM}. 

The aim of the present paper is to present, in a compact but systematic way, some arguments, in part already present in this discussion, and to propose a different physical interpretation. We assume a phenomenological point of view, without reference to any underlying fundamental microscopic theory. Since the relativistic kinematics has proven to be satisfactory, up to now, in laboratory experiments, we consider a ``deformation '' of the kinematic laws depending on a small continuous parameter, which many authors relate to the Planck length. 

Since in the present paper we are only interested in a phenomenological discussion of the propagation and the collision of particles, which are described in terms of classical observables, we do not consider quantum groups \cite{MR,LRZ}. Another possibility to introduce a new length or energy scale in a group-theoretical way (namely on the basis of symmetry principles) is to enlarge the symmetry group. The first attempt in this sense was Born's duality principle \cite{Born} and other possibilities have been considerd more recently \cite{Toller,Brandt}. These ideas have not been developed into complete theories.

Even if it is well known, for completeness we recall that the Lorentz group has a ``rigid`` structure of Lie group, namely a small deformation which is again a Lie group is isomorphic to the original group. It is sufficient to show that  the Lorentz Lie algebra is rigid, namely to prove that slightly modified structure coefficient, still satisfying the Jacobi identity and the antisymmetry requirement, define a Lie algebra isomorphic to the original one.  This rigidity property is shared by all the semisimple Lie algebras \cite{NR}, but not, for instance, by the Galilei Lie algebra, which can be deformed into the Lorentz Lie algebra, as it happens in the passage from classical to relativistic physics.

\section{Deformations of homogeneous spaces.}
Then we can only consider a possible deformation of the transformation properties of the four-momentum under the undeformed Lorentz group. We disregard the possibility that the deformed transformations involve other physical quantities, for instance the spin of the particle. We indicate by $\Pi$ the three-dimensional manifold, usually called the ``mass-shell'', defined in $R^4$ by the dispersion law (\ref{Disp}). The components of four-momentum can be considered as a (redundant) set of coordinates on $\Pi$. The Lorentz group acts continuously and transitively on $\Pi$, namely $\Pi$ is an homogeneous space. 

A standard construction \cite{Chevalley} shows that an homogeneous space is isomorphic to a quotient space ${\cal L}/{\cal H}$, where ${\cal L}$ is, in our case, the proper orthochronous Lorentz group and ${\cal H}$ is a closed subroup of ${\cal L}$, namely the stability subgroup of an arbitrary point of $\Pi$. Since the Lorentz group has dimension six and $\Pi$ has dimension three, the subgroup $\cal H$ must have dimension three. 

In the usual relativistic theory, $\Pi$ is an orbit in four-momentum space and $\cal H$ is one of the Wigner little groups \cite{Wigner}, namely $SO(2) = {\cal H}_1$ for massive particles and $E(2) = {\cal H}_0$ for massless particles. The other three-dimensional little group $SO(1, 2)$ is excluded because in the corrsponding orbit the energy has not a lower bound. The homogeneous spaces relevant for the description of particles, which we call $\Pi_1 = {\cal L}/{\cal H}_1$ and $\Pi_0 = {\cal L}/{\cal H}_0$, are defined in terms of the components of a four-vector $\pi^i$ by  the two equations
\begin{equation} \label{Homo1}
\pi_i \pi^i = 1, \qquad \pi^0 \geq 1,
\end{equation}
\begin{equation} \label{Homo0}
\pi_i \pi^i = 0, \qquad \pi^0 > 0.
\end{equation}
Note that the quantities $\pi^i$, which transform linearly under the Lorentz group, form a (redundant) set of coordinates on $\Pi$, but we do not identifiy them with the components of the four-momentum.

A deformation of a homogeneous space corresponds to a deformation of the corresponding stability group. This is not a group deformation of the kind described above: the deformed subgroup must be a closed Lie subgroup of $\cal L$ not conjugate to the original one, since conjugate subgroups define isomorphic homogeneous spaces. We recall that two subgroups $\cal H$ and  ${\cal H}'$  are conjugate if ${\cal H}' = g{\cal H}g^{-1}$, where $g \in {\cal L}$. We are not assuming that the deformed homogeneous space is an orbit in four-momentum space.

The deformations of the Lie subgroup $\cal H$ correspond to deformations of its Lie subalgebra $L({\cal H})$. This means to change slightly the three elements which form a basis of $L({\cal H})$ in such a way that they form a basis of a new subalgebra not conjugate to the original one. We say that two subalgebras are conjugate  if one is transformed into the other by a linear transformation of $L({\cal L})$ belonging to the adjoint representation of $\cal L$. 

Note that the deformation of a Lie subalgebra up to conjugation is not the same problem as the deformation of a Lie algebra up to isomorphism. They  are controlled by two different cohomology groups, denoted, respectively, by $H^1(L({\cal H}),L({\cal L})/L({\cal H}))$ and $H^2(L({\cal H}), L({\cal H}))$. If the Lie algebra $L({\cal H})$ is semisimple, the Whitehead lemma \cite{CeE} assures that both the cohomology groups are trivial and both kinds of deformation are forbidden.

This deformation problem can be solved by means of the powerful methods of homological algebra or, as we have done, by means of direct calculations, to long to be reported here.  The result is that the semisimple subalgebra $L({\cal H}_1) = o(3)$, in agreement with the general rule, cannot be deformed, while $L({\cal H}_0)$, which is not semisimple,  has deformations depending on two parameters. 

We introduce a basis in $L({\cal L}) = o(1, 3) = sl(2, C)$, composed of the generators of the rotations $M_r$ and the generators of the Lorentz boosts $L_r$ ($r = 1, 2, 3)$, which satisfy the familiar commutation relations. A basis of the Lie subalgebras obtained by deforming  $L({\cal H}_0)$ is given by
\begin{displaymath} 
A_1 = (1 + \mu) M_1 + (1 - \mu) L_2, \qquad
A_2 = (1 + \mu) M_2 - (1 - \mu) L_1, 
\end{displaymath}
\begin{equation}
A_3 = M_3 + \nu L_3.
\end{equation}
These are solutions of a linearized problem, which disregards higher order terms in the infinitesimal parameters $\mu$ and $\nu$. The existence of one of these solutions is only a necessary condition for the existence of a deformation. A further analysis shows that higher order corrections are possible only if $\mu \nu = 0$. If this condition is satisfied, there is no need of corrections, because, as one can easily see, the formula given above defines a Lie subalgebra for any finite value of the non-vanishing parameter.

For $\mu = \nu = 0$, we have the Lie algebra $L({\cal H}_0)$ of the Wigner little group $E(2)$. For $\nu = 0$, we obtain a Lie algebra conjugate to $o(3)$, if $\mu > 0$, and to $o(1,2)$, if $\mu < 0$. In the first case, this simply means to give a small mass to a massless particle, and we have already recalled that the orbit corresponding to the little group $SO(1, 2)$ is not physically acceptable.  

For $\mu = 0$, and any real value of $\nu$, the subgroup generated by this Lie subalgebra in the group $SL(2, C)$, locally isomorphic to $\cal L$, is composed of the matrices of the form
\begin{equation}
\left( \begin{array}{cc}
\exp(-it + \nu t) & z \\ 0 & \exp(it - \nu t)
\end{array} \right),
\end{equation}
where $t$ is real and $z$ is complex. For $\nu \neq 0$, an arbitrary element of $SL(2, C)$ can be uniquely decomposed into the product of an element of $SU(2)$ and an element of the group described by this formula. It follows that the corresponding homogeneous space is diffeomorphic to the manifold of $SU(2)$ and to the sphere $S_3$. If we consider the Lorentz group instead of $SL(2, C)$, the deformed homogeneous space is diffeomorphic to the manifold of the group $SO(3)$, to the sphere $S_3$ with the opposite points identified and to the three-dimensional real projective space. Note that for different values of $\nu \neq 0$, we obtain homogeneous spaces which have the same manifold structure (they are diffeomorphic), but they are not isomorphic as homogeneous spaces, since the Lie algebras of the corresponding stability subgroups are not conjugate. 

These deformed homogeneous spaces are compact manifolds and this could be a good property from the point of view of refs.\ \cite{AC4,MS}, since all the continuous coordinates are necessarily bounded. However, the rotation group acts transitively on them and no scalar function, representing the energy, can be defined. Moreover, an arbitrary change of the value of the energy can be obtained by means of a rotation. We conclude that these homogeneous spaces are pure mathematical curiosities and we do not consider them in the following. Note that the deformations we are considering are small locally, but they can change the global topology of $\Pi$, as it also happens in the deformation of $\Pi_0$ into $\Pi_1$.

The analysis summarized above may seem complicated, but the conclusion is simple: from the abstract point of view, namely up to isomorphisms, for the description of particles, we can use only the two famliar undeformed homogeneous spaces $\Pi_1$ and $\Pi_0$. We have, however, a wide freedom in choosing the components $p^i$ of the four-momentum as functions defined on $\Pi$. If we require, as all the authors involved in the discussion, that, under the rotation group, the energy behaves as a scalar and the momentum as a vector, we have to put
\begin{equation} \label{EnMom}
p^0 = f(\pi^0), \qquad  p^r = g(\pi^0) \pi^r, \qquad r = 1, 2, 3.
\end{equation}
We assume that the functions $f$ and $g$ are continuous functions of $\pi^0$ defined in the whole range of this variable. We also assume that $f$ and $|\vec\pi| g$ are increasing functions of $\pi^0$, so that each of the four scalar quantities $\pi^0$, $|\vec\pi|$, $p^0$ and $|\vec p|$ determines the other three uniquely.

It is possible that a particular choice of the coordinates is preferable in the framework of some fundamental theory, but it is clear that, from a phenomenological point of view, the physical phenomena can be described, in a more or less complicated way, with any choice of the coordinates on $\Pi$ and this choice is a matter of convenience. A natural requirement, however, is that the quantities $p^i$ represent the energy and the momentum which are really measured in the experiments. 

Note that the ultra-high-energy cosmic rays are not detected directly and their momentum cannot be measured from their deviation in a magnet. They interact with the nuclei in the high atmosphere and originate a shower of particles which are detected at the ground.  The energy attributed to the primary particle depends on a theory of the air shower and, in particular, on the assumption that the energy is conserved in the atmospheric collisions.  For this and other reasons, it would be highly desirable to choose the coordinates $p^i$ in such a way that they are conserved in collisions, but it is not evident that this is possible when the center-of-mass energy is very large.

\section{Conservation of energy-momentum.}
It is known that, if energy and momentum are conserved in the elastic collisions of identical particles, the eq.\ (\ref{EnMom}) must take the simple form 
\begin{equation} \label{EnMom1}
p^i = g \pi^i, \qquad i = 0,\ldots, 3,
\end{equation}
where g is a constant and an appropriate choice of the units and of the additive constant of the energy has been adopted. For massive particles, the argument is very old \cite{LT} and it is reported, in slightly different versions, in many textbooks \cite{Moller,Jackson}, but the interpretation is different, because in this old treatment the quantities $\pi^i$ play the role of the components of the four-velocity (actually, the argument is presented in terms of the three-dimensional velocity vector). However the only assumption used in the proof is that the quantities $\pi^i$ transform as the components of a four-vector and the argument can be used for our purposes. The proof is easily extended to massless particles.

In the laboratory experiments, energy and momentum are conserved with a high accuracy and therefore they are given by eq.\ (\ref{EnMom1}). If we consider only elastic scattering of identical particles, the value of the coefficient $g$ is undetermined. It is a remarkable experimental fact that the ratios of the quantities $g$ for the various massive particles can be determined consistently by requiring the conservation of four-momentum in many different low-energy collision processes. These quantities are identified with the masses.

The coordinates $p^i$, defined on $\Pi$ in this way, have the usual linear transformation properties and satisfy the usual dispersion law for any energy. We call them the components of the {\it linear} four-momentum. They are conserved at laboratory energies, but the conservation laws may not be valid at higher center-of-mass energies. Then, however, no other set of coordinates can satisfy the conservation laws and one should have other good reasons to prefer non-linear coordinates.

\section{The GZK cutoff.}
The collision of an ultra-high-energy proton with a CMB photon, when described in the centre of mass system, is a low energy process and, if we assume the relativity principle, the linear four-momentum is conserved, also in a terrestrial frame, and the cross-section too is the one measured in laboratory. We conclude that, if there are no preferred inertial frames, the GZK cutoff is valid for the linear energy. One can introduce non-linear coordinates from the beginning, with their anomalous dispersion law, but it is simpler to perform the calculations with the linear four-momentum and to transform the coordinates at the end. The result should be the same.

However, the collision of the proton with the atmospheric nuclei, even in the center-of-mass system, involves an unexplored energy range and the energy momentum could not be conserved. In fact, if a proton with energy above the GZK cutoff collides with a proton at rest, the center-of-mass energy is larger than $3 \times 10^5 \, GeV$, two orders of magnitude more than the energy attainable with any reasonable terrestrial collider.

If we want to explain the cosmic ray anomalies without abandoning the relativity principle, we have to assume that the measured energy is larger than the linear energy, due to a failure of the conservation laws in the atmospheric proton-nucleus interactions. This seems to be the only alternative to the introduction of privileged inertial frames. On the other hand, if future more precise observations show that the GZK cutoff is respected \cite{BW} we shall obtain an argument in favour of the energy-momentum conservation in the ultra-high energy hadron collisions.  We stress that, if one is disposed to abandon the Lorentz symmetry and the relativity principle, the situation is completly different and a violation of the GZK cutoff may well be consistent  with the validity of the conservation laws at ultra-high energies.

In these considerations we, as well as other authors, have implicitly assumed that the four-momentum is conserved in the propagation of particles in vacuum. Note that this assumption does not depend on the choice of the coordinates in $\Pi$. If the particle travels along a distance not negligible with respect to the cosmological scale, one cannot consider a global inertial frame and it is natural to assume the conservation of the four-momentum measured in a local frame parallel transported along the world line of the particle. This effect, as is well known, decreases the energy of the particle measured in a terrestrial frame and makes the GZK cutoff more stringent. 

With respect to other proposals, the point of view discussed in this paper shifts the attention from the proton-gamma interaction in the intergalactic space to the proton-nucleus interaction in the high atmosphere. This is reasonable, since, even assuming the relativity principle, no information on the second process is available without a considerable extrapolation of our experimental knowledge. The attention is also shifted from a deformation of the dispersion law to a deformation of the conservation law. 

In a recent article \cite{AC3}, what we call the linear four-momentum is introduced (as an auxiliary concept) and an explicit relation with the adopted definition of energy is discussed. The formula proposed (up to higher order terms), with our notations, takes the form
\begin{equation}
m \pi^0 = p^0 - \frac{(p^0)^3}{m E_p}, \qquad
p^0 < \left(\frac{m E_p}{3}\right)^{1/2}, 
\end{equation}
where $m$ is the mass and $E_p$ defines the new energy scale. We have added an inequality that assures that the relation is monotonic and invertible. 

From the point of view of the present paper, this can be interpreted as the relation  between the linear energy $m \pi^0$, which satisfies the GZK cutoff, and the actually measured energy $p^0$. The measured energy can be larger than the GZK cutoff, but at most by a factor $3/2$, which does not seem to be sufficient. However, this formula is just an example of a general class of relations, in which the various energy scales are combined in a carefully chosen way, namely
\begin{equation}
m \pi^0 = (m E_p)^{1/2} F\left((m E_p)^{-1/2} p^0 \right).
\end{equation}
If $E_p$ is the Planck energy, for a reasonable choice of the increasing function $F$, the discrepancy with the GZK cutoff is avoided. From the point of view we are considering, this formula should be derived from the theory of the air shower on the basis of a specific assumption about the deformation of the four-momentum conservation.

We remark that the non-conservation of four-momentum imposes much harder problems on the general ideas of theoretical physics than the existence of privileged inertial frames and the consequent breaking of Lorentz symmetry. While in the second case it is sufficient to introduce a vector or tensor field with a non vanishing vacuum expectation value, in the first case one has to modify the familiar connection between symmetry and consevation laws or to find a breaking mechanism of the symmetry with respect to spacetime translations, effective during the very short interaction time of a collision. It is clear that one has to find new ideas, probably in the field of quantum gravity. The change of point of view we are suggesting does not affect the idea \cite{ACEMNS,Kifune,ACP,AC1} that there may be a connection between quantum gravity and the physics of ultra-high energy cosmic rays.

\end{document}